\newcount\Comments  
\documentclass[prl, reprint, superscriptaddress]{revtex4-1}
\usepackage[english]{babel}
\usepackage[utf8]{inputenc}
\usepackage{siunitx}
\usepackage[colorinlistoftodos, color=green!40, prependcaption]{todonotes}
\usepackage{graphicx}
\usepackage[colorinlistoftodos]{todonotes}
\usepackage{amsthm}
\usepackage{mathtools}
\usepackage{physics}
\usepackage{xcolor}
\usepackage{graphicx}
\usepackage[left=23mm,right=13mm,top=35mm,columnsep=15pt]{geometry} 
\usepackage{adjustbox}
\usepackage{placeins}
\usepackage[T1]{fontenc}
\usepackage{lipsum}
\usepackage{csquotes}
\usepackage[pdftex, pdftitle={Article}, pdfauthor={Author}]{hyperref} 
\usepackage{fancyhdr}
\usepackage{xr}
\makeatletter

\newcommand*{\addFileDependency}[1]{
\typeout{(#1)}
\@addtofilelist{#1}
\IfFileExists{#1}{}{\typeout{No file #1.}}
}\makeatother

\newcommand*{\myexternaldocument}[1]{%
\externaldocument{#1}%
\addFileDependency{#1.tex}%
\addFileDependency{#1.aux}%
}

\myexternaldocument{supplement}

\newcommand{\kibitz}[2]{\ifnum\Comments=1\textcolor{#1}{#2}\fi}

\usepackage{natbib}
\begin{document}

\title{Ultrafast Photo-induced Phase Change in SnSe}

\author{Benjamin J. Dringoli}
    \affiliation{Department of Physics, McGill University, Montreal, QC, Canada H3A2T8}

\author{Mark Sutton}
    \affiliation{Department of Physics, McGill University, Montreal, QC, Canada H3A2T8}
    
\author{Zhongzhen Luo}
    \affiliation{Department of Chemistry, Northwestern University, Evanston, Illinois 60208, United States}
    \affiliation{Key Laboratory of Eco-Materials Advanced Technology, College of Materials Science and Engineering, Fuzhou University, Fuzhou 350108, P. R. China}

\author{Mercouri G. Kanatzidis}
    \affiliation{Department of Chemistry, Northwestern University, Evanston, Illinois 60208, United States}

\author{David G. Cooke}
    \email[Correspondence email address: ] {david.cooke2@mcgill.ca}
    \affiliation{Department of Physics, McGill University, Montreal, QC, Canada H3A2T8}


\begin{abstract}

\noindent Time-resolved multi-terahertz (THz) spectroscopy is used to observe an ultrafast, non-thermal electronic phase change in SnSe driven by interband photoexcitation with 1.55 eV pump photons. The transient THz photoconductivity spectrum is found to be Lorentzian-like, indicating charge localization and phase segregation. The rise of photoconductivity is bimodal in nature, with both a fast and slow component due to excitation into multiple bands and subsequent intervalley scattering. The THz conductivity magnitude, dynamics, and spectra show a drastic change in character at a critical excitation fluence of approximately $6$~mJ/cm$^2$ due to a photo-induced phase segregation and a macroscopic collapse of the band gap.

\end{abstract}


\maketitle

\begin{figure}[t!]
    \centering
    \includegraphics[width=1\columnwidth]{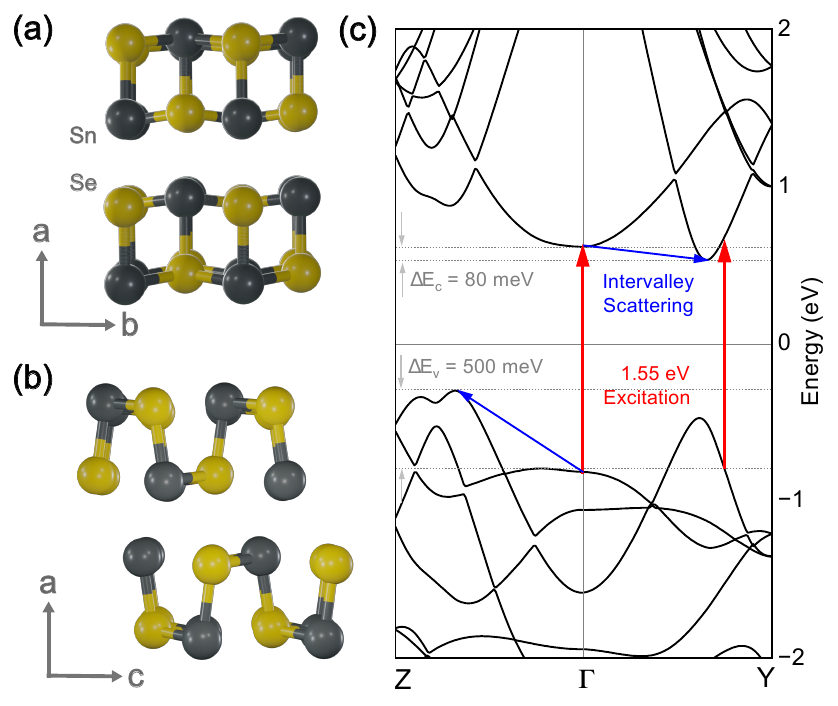}
    \caption{(a,b) SnSe layered atomic structure along the in-plane zigzag b-axis and armchair c-axis in the room temperature \textit{Pnma} phase (c) \textit{Pnma}-phase SnSe electronic band structure from \textit{ab initio} calculations using data provided from Ref.~\cite{TayariPRB2018} with permission from the authors. Red arrows show carrier excitation from the 1.55 eV pump, blue arrows show post-excitation relaxation pathways to band extrema.}
    \label{fig1}
\end{figure}

\indent Tin Selenide (SnSe) is a quasi-two dimensional semiconductor exhibiting record-setting thermoelectric properties, with a figure-of-merit $ZT = 2.62$ along the in-plane $b$-axis \cite{ZhaoNAT2014, GuoPRB2015,ChangCM2018,AlsalamaRAMS2020,TayariPRB2018}. 
A polar analogue of black phosphorous, at room temperature SnSe has a layered, orthorhombic $Pnma$ phase with zig-zag and armchair ordering shown in Fig.~\ref{fig1}(a) and (b), respectively. In the $Pnma$ phase, it possesses an indirect band gap of 0.9~eV \cite{PatelEPJB2019} and a direct gap of 1.21~eV \cite{DangPSSA1984}, shown in Fig.~\ref{fig1}(c). At $T_c = 807$~K, a $Pnma \rightarrow Cmcm$ structural phase transition occurs, reducing thermal conductivity due to phonon-phonon scattering arising from strong anharmonicity. The origin of this transition has been intensely studied in the past several years, being attributed to either a displacive transition typically seen in ferroelectrics or a dynamic order-disorder transition \cite{SkeltonPRL2016, BansalPRB2016, TangACS2018, ChangMTP2018, LiuPRB2018, HongMTP2019, WuNJP2020, LuPRB2021,JiangNCOM2023}. The displacive transition is primarily driven by atomic motion along the c-axis \cite{ChattopadhyayJPCS1986}, accompanied by phonon softening and an eventual collapse of terahertz (THz) frequency zone center optical phonons \cite{ChattopadhyayJPCS1986, AseginolazaPRL2019, LangianNC2020}. Very recent neutron scattering measurements, however, support a disordered high temperature phase with local contributions from the $Pnma$ phase on length scales of a few unit cells \cite{JiangNCOM2023}. Phonon anharmonicities in SnSe are driven by electronic bonding instabilities involving Se 4p-orbitals and Sn 5s-orbitals containing lone pair electrons \cite{ZhaoEES2016,XiaoPRB2016,KangNL2019,LiNP2015,HongMTP2019,MaPRB2018,LangianNC2020}, whose states comprise the photoacessible $\Gamma$-valley valence band. Optical excitation can selectively perturb these deeper lying bands and reveal novel, thermally inaccessible phases.
\\
\indent Ultrafast scattering and spectroscopy can disentangle the complexity underlying these dynamic lattice instabilities \cite{LanSCI2019,OttoPNAS2019,LiuNAT2012,FaustiSCI2011}, leading to a better understanding of the microscopic interactions underlying thermoelectric properties. Recently, structural probes such as ultrafast electron diffraction (UED)\cite{WangNPJQM2021, ReneDeCotretPNAS2022}, time-resolved x-ray diffraction (TR-XRD) \cite{HuangPRX2022, HuangARX2023}, and transient reflectivity (TR) measurements \cite{HanJPCL2022} have revealed novel, non-thermal structural dynamics following femtosecond excitation at photon energies of 1.55 eV ($\lambda_p = 800$~nm). These have indicated sub-picosecond transitions towards the $Cmcm$ phase driven by displacive excitation of $A_g$ phonons \cite{HanJPCL2022}, domain formation mediated by photoinduced interlayer strain fields \cite{WangNPJQM2021}, bimodal polaron formation \cite{ReneDeCotretPNAS2022}, and transient atomic motions towards a thermally inaccessible $Immm$ phase \cite{HuangPRX2022,HuangARX2023}. Given the diversity of structural phenomena indicated by these prior works, a complementary survey of the electronic behavior under these excitation conditions is needed.
\\
\indent In this work, we use ultra-broadband time-resolved terahertz (THz) spectroscopy to probe the electronic response of SnSe under identical excitation conditions as these previous ultrafast structural probe measurements ($\lambda_p = 800$~nm, $F=0.1 - 13.2$ mJ/cm$^2$). We observe non-Drude THz conductivity spectra exhibiting bimodal formation dynamics consistent with previous UED work reporting polaron formation after ultrafast excitation \cite{ReneDeCotretPNAS2022}. Moreover, we find evidence for an electronic phase transition occurring at a critical pump excitation fluence $F_c \approx 6$~mJ/cm$^2$, where THz conductivity magnitude, dynamics and spectra change sharply. This excitation density is consistent with the proposed $Pnma\rightarrow Immm$ destabilization threshold \cite{HuangPRX2022}.


\indent Time-resolved THz spectroscopy (TRTS) was performed using single-cycle, multi-THz pulses shown in Supplemental Fig.~1(a). These pulses were generated by a two-color laser plasma in dry air, formed by co-focusing 35~fs-duration fundamental (800~nm) and second harmonic (400~nm) pulses from a 1~kHz amplified Ti:Sapphire femtosecond laser system \cite{DaiPRL2006,CookOL2000,KimJQE2012,ThomsonOE2010,JepsenPRL2011}. These multi-THz pulses contain Fourier components continuously covering a $1 - 20$~THz spectral range ($4-85$~meV), as shown in Supplemental Fig.~1(b). A schematic of the spectrometer is shown in Supplemental Fig.~2 and further details on the setup are provided in Ref.~\cite{LanSCI2019}. The SnSe sample was a 3~mm x 5~mm x 3.6~mm, high quality single crystal of SnSe grown using methods as described in Ref.~\cite{ZhaoNAT2014}, and from the same parent crystal as in Ref.~\cite{ReneDeCotretPNAS2022}. The crystal was measured via reflection at normal incidence with the THz and pump pulses linearly polarized along the b-axis. Other crystal orientations were measured, with results shown in Supplemental Fig.~3, however no significant change in response was observed.

\indent The reflected THz field was directly sampled in time $t$ using air-biased coherent detection at various pump-probe delay times $\tau$ \cite{DaiPRL2006,HoOE2012}. Both the reflected field in the presence of a pump pulse, $E_p(t,\tau)$, and the differential $\Delta E(t,\tau) = E_p(t,\tau) - E_r(t)$ are measured simultaneously in a double-modulation scheme \cite{IwaszczukOE2009}, where $E_r(t)$ is the reference THz field reflected from the unexcited sample. The pulse delays are implemented such that all points in the measured THz fields have experienced a constant $\tau$, so a Fourier transformation along the real time axis $t$ is devoid of any convolution artifacts and directly provides a temporal snapshot of the frequency-dependent response \cite{NemecJCP2002}. Intrinsic artifacts that can appear in TRTS, as mentioned in Ref.~\cite{OrensteinPRB2015}, are not present (see Supplemental Fig.~4) as carrier momentum relaxation times are extremely short \cite{ChavesPCCP2021}. The pump-induced differential reflectivity $\Delta \Tilde{r}(\omega,\tau)/r_0(\omega) = \Delta E(\omega,\tau) / E_{r} (\omega)-1$, where $r_0(\omega)$ is the static reflectivity, was subsequently extracted by Fourier analysis. The static $r_0(\omega)$ is dominated by a single reststrahlen band for THz polarization parallel to the b-axis \cite{EfthimiopoulosPCCP2019}, shown in Supplemental Fig.~1, which prohibits the extraction of the optical conductivity below ${\sim}5$~THz. The THz reflectivity is finally related to the pump-induced differential optical conductivity $\Delta \Tilde{\sigma}(\omega,\tau) = \Delta\sigma_1(\omega,\tau)+i\Delta\sigma_2(\omega,\tau)$ by modelling the photoexcited layer as a slab by \cite{HegmannSPIE2002}:
\\
\begin{equation}
   \Delta \Tilde{\sigma}(\omega,\tau) = \frac{1}{Z_0 d}  \frac{(\Delta \Tilde{r}/r_0)(\Tilde{n}^2-1)}{(\Delta \Tilde{r}/r_0)(1-\Tilde{n})+2},
\end{equation}
\\
where $d = 60$~nm is the pump penetration depth at $\lambda = 800$~nm \cite{MakinistianPSSB2009}, $Z_0 = 377~\Omega$ is the impedance of free space, and $\Tilde{n}(\omega)$ is the static complex THz refractive index \cite{EfthimiopoulosPCCP2019}. Despite reports of saturable absorption in SnSe nanosheets at much lower peak powers \cite{WangIPT2020}, the pump absorption was found to be independent of excitation fluence and average heating of the sample surface was estimated by infrared imaging to be $<15^\circ$C, shown in Supplemental Figs.~5 and 6, respectively.

\begin{figure}[t]
    \centering
    \includegraphics[width=1.0\columnwidth]{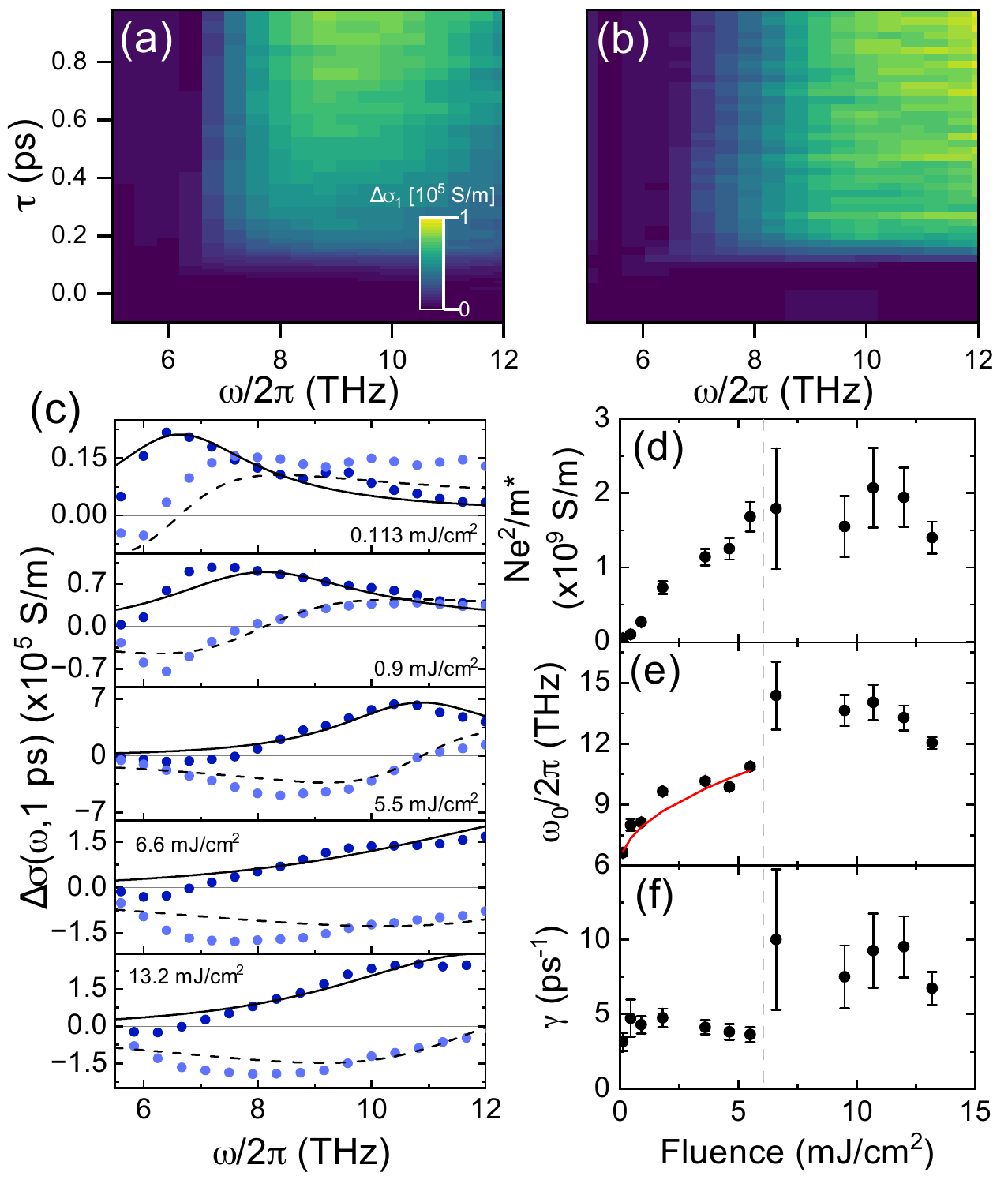}
    \caption{(a,b) Two-dimensional real conductivity $\Delta\sigma_1(\omega,\tau)$ maps for 1.8 and 12.0~mJ/cm$^2$. (c) Complex THz differential photoconductivity spectra $\Delta \Tilde{\sigma} (\omega,\tau) = \sigma_1 (\omega,\tau) + i \sigma_2 (\omega,\tau)$ for various fluences. $\Delta\sigma_1(\omega, \tau=$ 1~ps) shown in dark colors, $\Delta\sigma_2(\omega,\tau=$ 1~ps) shown in light colors, and Lorentzian fits shown in black solid and dashed lines. Note the significant difference in vertical axis scales. Fluence dependence of the (d) Lorentzian oscillator strength $Ne^2/m^*$, (e) center frequency $\omega_0$, and (f) linewidth $\gamma$. $\omega_0$ is fitted with a scaled $\sqrt{F}$ function, shown in red and discussed in the text.}
    \label{fig2}
\end{figure}

\begin{figure}[t!]
    \centering
    \includegraphics[width=\columnwidth,trim={1cm 0cm 0.5cm 0cm},clip]{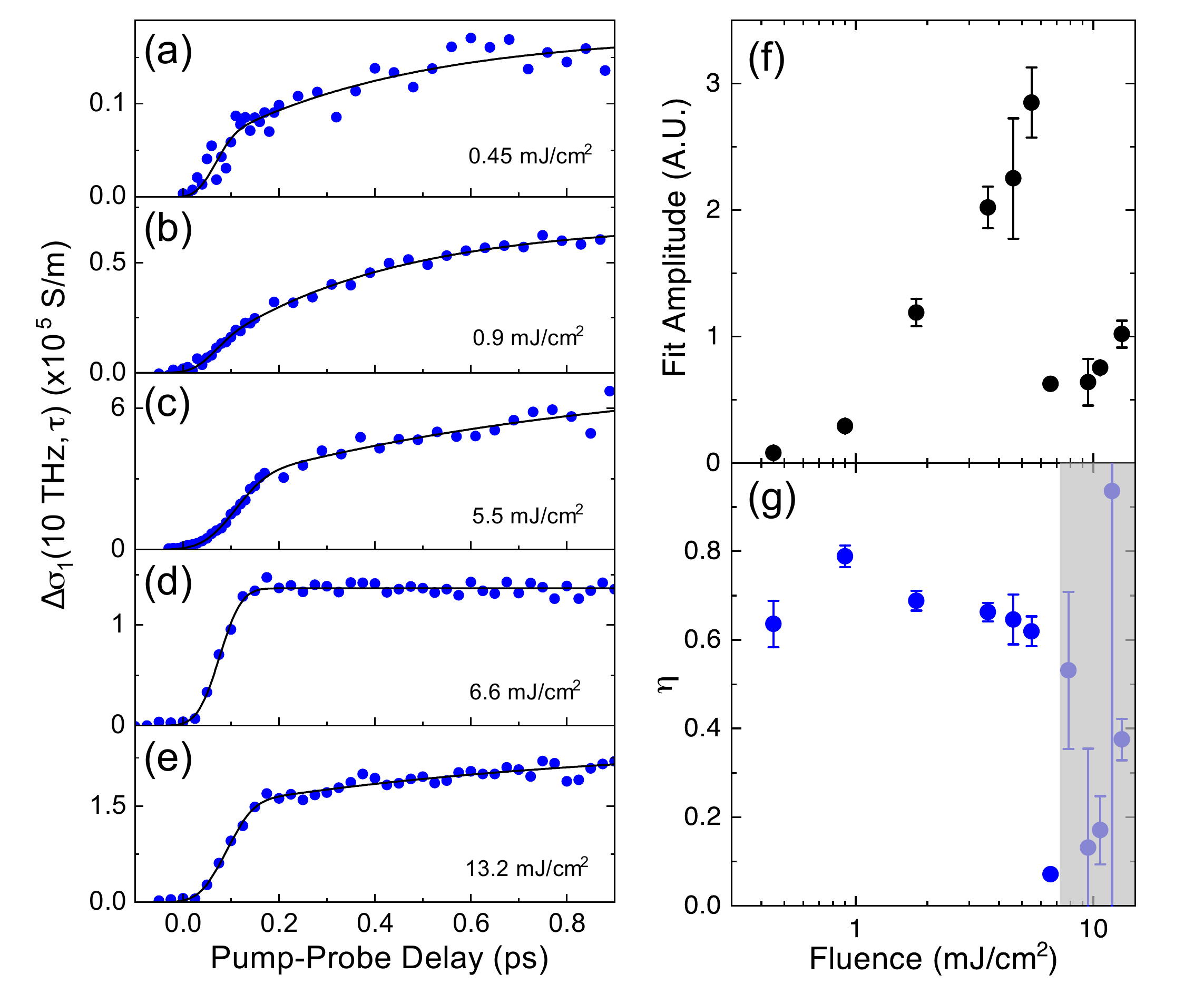}
    \caption{(a-e) $\Delta\sigma_1(\omega,\tau)$ dynamics at $\omega/2\pi = 10$~THz for indicated pump fluences $F$. Rate equation model fits (Eqn.~3) are shown as black lines. Note the differences in vertical axis scaling. (f) $\Delta \sigma_1$(10~THz, 1~ps) as a function of pump fluence, showing approximately linear scaling below $F_c$ and a sharp drop with little scaling above $F_c$. (g) Fitted proportion of carriers deposited in band extrema $\eta$ (Eqn. 3) as a function of fluence showing sharp drop near $F_c$.}
    \label{fig3}
\end{figure}

\indent Two-dimensional $\Delta \sigma_1 (\omega,\tau)$ maps for frequencies above all phonon branches ($\omega/2\pi>5.5$~THz), where the electronic intraband response dominates, are shown in Fig.~\ref{fig2} for pump fluences of (a) 1.8~mJ/cm$^2$ and (b) 12.0~mJ/cm$^2$. Additional 2D maps for varying fluences are presented in Supplemental Fig.~7. In both low and high fluence regimes, the conductivity has Lorentzian character despite all assumptions of SnSe being a band semiconductor where the expectation is a simple Drude response. Such a broadened Lorentzian response can arise from phase heterogeneity and the formation of domains that favor conduction over smaller length scales (higher frequencies), which has been observed in SnSe at these fluences \cite{WangNPJQM2021}. The optical conductivity of a heterogeneous phase system is usually described by effective medium theories (e.g. Bruggeman), and typically displays a monotonically increasing $\sigma_1(\omega)$ from zero frequency \cite{KuzelJPD2014}. However, here $\sigma_1(\omega)$ exhibits a clear onset at the LO phonon energy of $\sim20$~meV ($\sim5$~THz). This could indicate that photoexcited carriers are polarons \cite{SioNATPHYS2023}, as suggested by calculations showing strong electron-phonon coupling near $\Gamma$ \cite{CarusoPRB2019} and previous UED measurements indicating the formation of a 3D polaron with radius $\sim4.2$ {\AA} after photoexcitation \cite{ReneDeCotretPNAS2022}.
\\
\indent To quantify the evolution of spectral shape, we fit the data to a Lorentzian response given by
\begin{equation}
    \Delta \Tilde{\sigma}(\omega) = \frac{N e^2}{m^*} \frac{\omega}{i(\omega_0^2-\omega^2)+\omega\gamma}
    \label{lorentzian}
\end{equation}
with $Ne^2/m^*$ as the oscillator strength, $\omega_0$ the resonance frequency, and $\gamma$ the scattering rate. This simple phenomenological model fits both $\sigma_1$ and $\sigma_2$ reasonably well over the entire fluence regime for all times, as shown in Fig.~\ref{fig2}(c) and in Supplemental Fig.~8. The fit parameters $Ne^2/m^*$, $\omega_0$, and $\gamma$ for all fluences are shown in Fig.~\ref{fig2}(d),(e) and (f), respectively. Two fluence regimes are identified: a low ($F< 6.6$~mJ/cm$^2$) and a high ($F\geq 6.6$~mJ/cm$^2$) fluence regime. In the low fluence regime, $Ne^2/m^*$ increases linearly and monotonically with fluence, as expected for unsaturated interband excitation of charge carriers. The scattering rate $\gamma$ remains relatively constant in this fluence range, reflecting consistency of the absorption pathways shown in Fig.~\ref{fig1}(c) for these fluences. The fitted $\omega_0$ scales with an approximately $\sqrt{F}$ dependence just above LO phonon energy. Localization of charges is implied in a Lorentzian response, however the increase of $\omega_0$ with fluence is inconsistent with models of weak localization \cite{LeePRB1993} or back-scattering \cite{SmithPRB2001}. Instead, this implies a plasmonic origin where such a $\omega_0\propto\sqrt{F}$ dependence is expected for a heterogeneous medium, provided there is isolation of neighboring conducting domains such that depolarization fields contribute \cite{KuzelJPD2014,JoyceSST2016}. The long-time relaxation of the system, shown in Supplemental Fig.~9, occurs on a ${\sim}150$~ps time scale, consistent with previous measurements reporting interband recombination \cite{YanOM2020}. This relaxation is also seen in the redshifting of the spectral peak at long times shown in Supplemental Fig.~10, again consistent with a system of heterogeneous conductive media. At a critical fluence $F_c = 6.6$~mJ/cm$^2$, a discontinuous increase in $\omega_0$ and $\gamma$ occurs while $Ne^2/m^*$ saturates. This sudden change suggests a charge density-driven electronic phase transition occurring on ultrafast time scales.  
\\
\indent Further evidence for an electronic phase transition is seen in the dynamics of $\sigma_1(\omega,\tau)$, monitored at $\omega/2\pi = 10$~THz, far above all phonon branches and where $\Delta\sigma$ is dominated by electronic excitations. Fig.~\ref{fig3}(a-e) shows the clear bimodal dynamics in the sub-picosecond manifestation of photoconductivity. In the low fluence regime, there are two distinct components: a $40$~fs instrument response-limited rise after photoexcitation and a slower ${\sim}200$~fs rise. Examining the $Pnma$ band structure shown in Fig.~\ref{fig1}(c), we can classify pump absorption into two main channels: 1) excitation into the high effective mass, non-dispersive bands about the $\Gamma$-point which subsequently undergo interband scattering to the lower-mass band extrema; and 2) excitation of hot, mobile charge carriers distributed throughout the Brillouin zone that cool via intraband relaxation. The $\Gamma$ valley carriers have an effective mass of $1.20m_e$ \cite{GuoPRB2015}, eight times that of band extrema along $\Gamma - Y$ and $\Gamma - Z$ at $0.15m_e$ \cite{DasJPCM2020}. Therefore, the measured THz conductivity is dominated by $n_{free}$, with the slow rise representing the scattering of heavier $n_\Gamma$ carriers to the extrema, enabled through the high electron-phonon coupling near the $\Gamma$ point \cite{CarusoPRB2019}. The conductivity dynamics can therefore be well fit by a two-component rate equation model:
\begin{equation}
\begin{split}
    &\frac{d n_{\Gamma}(t)}{dt} = \eta \ \text{exp}[-(t/\sqrt{2} \tau_p)^2)] - n_{\Gamma}(t)  \gamma_{inter} \\
    &\frac{d n_{\text{free}}(t)}{dt} = (1-\eta)  \text{exp}[-(t/\sqrt{2} \tau_p)^2)] + n_{\Gamma}(t)  \gamma_{inter} \\
    &\Delta \sigma_1 (t) = A \left( \frac{e \tau}{m^*_f} n_{\text{free}}(t)+\frac{e \tau}{m^*_{\Gamma}} n_{\Gamma}(t)\right).
    \label{ratemodel}
\end{split}
\end{equation}
Here $n_{\Gamma}$,~$n_{\text{free}}$ are the populations in the $\Gamma$ and mobile bands, respectively, $\eta$ is the fraction of charge carriers excited into the $\Gamma$ valley, $\tau_p$ is the pump pulse duration, $e$ is the electron charge, $A$ is an amplitude scaling factor, and $m^*$ are the effective masses. As a simplifying assumption, the momentum scattering time $\tau$ is assumed to be 20~fs for both populations\cite{ChavesPCCP2021}. The intervalley scattering rate, $\gamma_{inter}\approx 1-2$~THz, is typical of polar semiconductors at room temperature \cite{ShahPRL1987}. When $F\geq6.6$ mJ/cm$^2$, the same fluence where the spectra sharply blue shifted, the magnitude of the response $A$ drops sharply as shown in Fig.~\ref{fig3}(f). At the same fluence, the slow scattering process and associated rise in conductivity, represented by $\eta$ and shown in Fig.~\ref{fig3}(g), vanishes and only recovers slightly at the higher fluence. This indicates the sudden elimination of intervalley scattering from the $\Gamma$-valley. Such a drastic change in the magnitude, spectra and dynamics of intraband conductivity at $F_c$ is a strong indication of a photo-induced, ultrafast change of macroscopic phase over THz wavelength length scales.

\indent A photo-induced $Pnma$-$Immm$ structural phase transition was recently observed in SnSe accompanied by a fluence-dependent softening of A$_{g}$ phonon modes \cite{HuangPRX2022}. This Peierls-like structural instability is driven by the removal of electrons in nondispersive electronic bands derived from both Se $4 px$- and Sn $5 s$-orbitals \cite{HuangARX2023}, which form the non-dispersive bands at $\Gamma$ \cite{HongMTP2019}. The authors estimated the $Pnma$ phase is fully destabilized in favor of the $Immm$ phase at a photoexcited carrier density of $0.2$ holes per formula unit, corresponding to $9\times10^{20}$~cm$^{-3}$. The corresponding critical density from these measurements is estimated as $n_c = F_c(1-R)\eta \lambda/h c d\approx 2\times 10^{21}$~cm$^{-3}$, in fair agreement. The transition to the $Immm$ phase is expected to cause band structure renormalization and a closure of the electronic band gap, leading to semimetallic electronic dispersion \cite{HuangPRX2022}. A transition to a semimetallic band structure could explain the fixed finite-frequency spectral peak above $F_c$ as a new interband THz absorption would appear in the mid-IR. Additional support for this idea is the relatively fluence-independent conductivity above $F_c$ shown in Fig.~\ref{fig2}(d), which would would be expected for a system whose gap is collapsed as excitation only serves to heat the charge distribution. While a recent time-resolved ARPES study of SnSe observed some evidence of spectral weight within the gap as early as 40 fs following photoexcitation, the 800~nm excitation fluence was significantly below $F_c$ at 0.95~mJ/cm\cite{OkawaSM2023}.


\indent In conclusion, time-resolved THz spectroscopy was used to probe the ultrafast electronic response of SnSe under fs optical excitation. The transient THz optical conductivity was Lorentzian in character whose resonant frequency is carrier density dependent. This suggests photocarrier localization into heterogeneous conducting domains and phase co-existence on the nanoscale, driven by optical excitation and the removal of bonding electrons. Bimodal photoconductivity dynamics were observed and could be explained by multi-valley excitation into both free and localized states related to the non-dispersive bands derived from the Sn 5s-Se 4p hybridization. At a critical fluence of $F_c\approx6$~mJ/cm$^2$, the spectra, dynamics, and magnitude of the THz conductivity change drastically indicating a macroscopic change of phase. The vanishing of intervalley scattering and the near constant conductivity magnitude with further increase of pump fluence suggests a collapse of the band gap, which is consistent with a recent study suggesting SnSe moves towards a thermally inaccessible $Immm$ phase. This motivates further high fluence TR-ARPES measurements and time-resolved spatial probes of nanoscale conductivity to potentially resolve this novel, transient phase of SnSe. 


\section{Acknowledgments}
BJD and DGC are grateful for the SnSe electronic band structure data from Ref. \cite{TayariPRB2018} provided by D. Rybkovskiy and for conversations with B. Siwick. This work was supported through funding by NSERC and Mitacs programs. The Northwestern University personnel and research work was supported by the Department of Energy, Office of Science, Basic Energy Sciences under Grant DE-SC0024256 (design and synthesis of thermoelectric materials).


\bibliography{SnSe_Mar23_Abbrv}

\end{document}


\title{Ultrafast Photo-induced Phase Change in SnSe:\\ Supplemental Information}

\author{Benjamin J. Dringoli}
    \affiliation{Department of Physics, McGill University, Montreal, QC, Canada H3A 2T8}

\author{Mark Sutton}
    \affiliation{Department of Physics, McGill University, Montreal, QC, Canada H3A 2T8}
    
\author{Zhongzhen Luo}
    \affiliation{Department of Chemistry, Northwestern University, Evanston, Illinois 60208, United States}
    \affiliation{School of Materials Science and Engineering, Fuzhou University, Fuzhou 350116, China}
    
\author{Mercouri G. Kanatzidis}
    \affiliation{Department of Chemistry, Northwestern University, Evanston, Illinois 60208, United States}

\author{David G. Cooke}
    \email[Correspondence email address] {david.cooke2@mcgill.ca}
    \affiliation{Department of Physics, McGill University, Montreal, QC, Canada H3A 2T8}

\date{\today} 

\keywords{thermoelectric, SnSe, polarons, electron-phonon coupling, conductivity, dynamics, terahertz, spectroscopy, photodoping, phase transition, pnma, immm}

\maketitle

Example time- and frequency-domain representations of the terahertz (THz) probe pulses used in this work are shown in Supplemental Fig. \ref{sfig4}(a) and (b), respectively. The b-axis of SnSe was the main focus of this study due to its in-plane orientation as well as its relatively simple static reflectivity dominated by a single TO phonon at $\omega_{TO} = 2.85$~THz [44]. This is shown by the measured change in reflectivity, shown in Supplemental Fig. \ref{sfig4}(c), which has a minimum in the high reflectivity region and an increase in reflected signal in the low reflectivity region after pumping.

\begin{figure}[h]
    \centering
    \includegraphics[width=0.75\columnwidth]{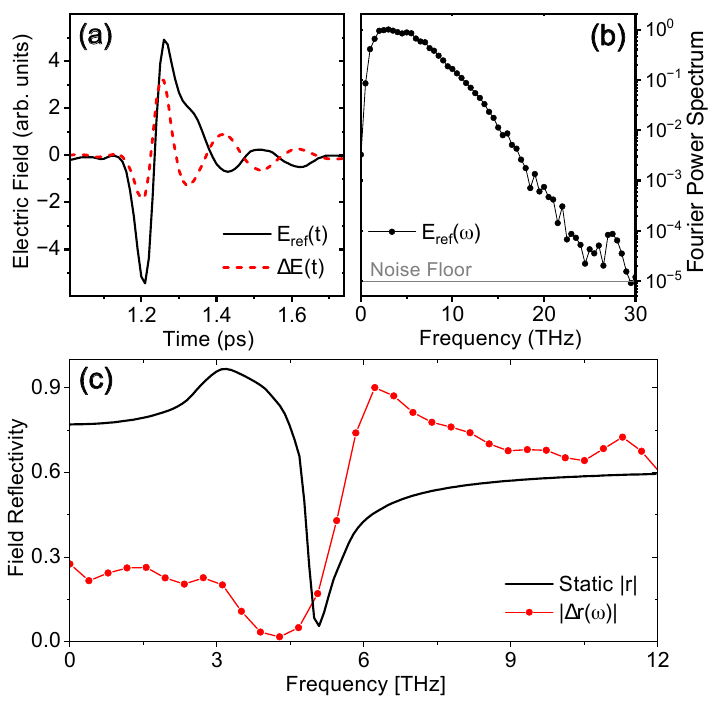}
    \caption{(a) Time-domain THz electric field traces showing the unpumped reference (black) and pump-induced change (red). (b) Fourier power spectrum of recorded reference pulse showing >20 THz bandwidth. (c) SnSe b-axis THz reflectivity (black) and measured change in reflectivity at 4.62 mJ/cm$^2$ and $\tau=1$~ps (red). Background reflectivity is taken from FTIR measurements in Efthimiopoulos et al.}
    \label{sfig4}
\end{figure}

\clearpage

Data for this experiment was taken using a THz spectrometer based on generation and detection of THz pulses in air, shown in Supplemental Fig. \ref{sfig1}. THz pulses generated from a two-color laser plasma are collected and refocused onto the sample using off-axis parabolic mirrors, with a silicon wafer allowing the pulse reflected from the sample to be collected for detection. An 800~nm pump pulse is introduced collinearly through a 2.5~mm hole in the focusing parabolic mirror with little effect on the THz pulse reflection due to the Laguerre-Gaussian mode of the two-color plasma emission [38]. The pump pulse is delayed with respect to the THz probe pulse using mechanical delay stages with a $\sim$7~fs step size. The reflected THz pulse is sampled using air-biased coherent detection [34,39], where the addition of a high-voltage local oscillator field allows for the imprinting of the instantaneous THz electric field onto the second harmonic of an 800~nm sampling pulse. The second harmonic signal is modulated by the local oscillator field at 500~Hz and is detected using an avalanche photodiode, with the modulated component recovered with a lock-in amplifier. This modulation scheme is also used with a mechanical chopper on the pump line to extract the pump-induced change in THz reflectivity.

\begin{figure}[h]
    \centering
    \includegraphics[width=0.8\columnwidth]{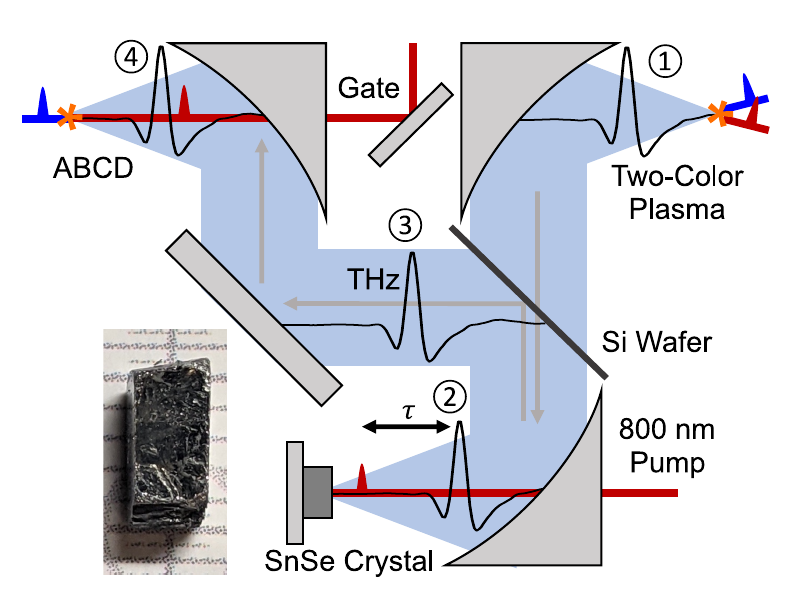}
    \caption{Schematic representation of the ultrabroadband THz spectrometer used for these measurements. THz pulses are generated in a two-color laser plasma, reflected off the sample surface, and detected using Air-Biased Coherent Detection (ABCD). This all-air experiment scheme removes bandwidth limitations from using solid state crystals with non-negligible absorption and dispersion.}
    \label{sfig1}
\end{figure}

\clearpage

Multi-fluence data was primarily taken with the THz electric field parallel to the SnSe b-axis (zigzag direction), but additional scans were taken with field aligned along the c- and a-axes (armchair and stacking directions, respectively). These are shown in Supplemental Figure \ref{sfig2}. All three crystal axes showed similar conductivity spectra and dynamics above all phonon energies after photoexcitation with 800~nm photons, showing the isotropic nature of the photoinduced effects in SnSe at early times.

\begin{figure}[h]
    \centering
    \includegraphics[width=0.85\columnwidth]{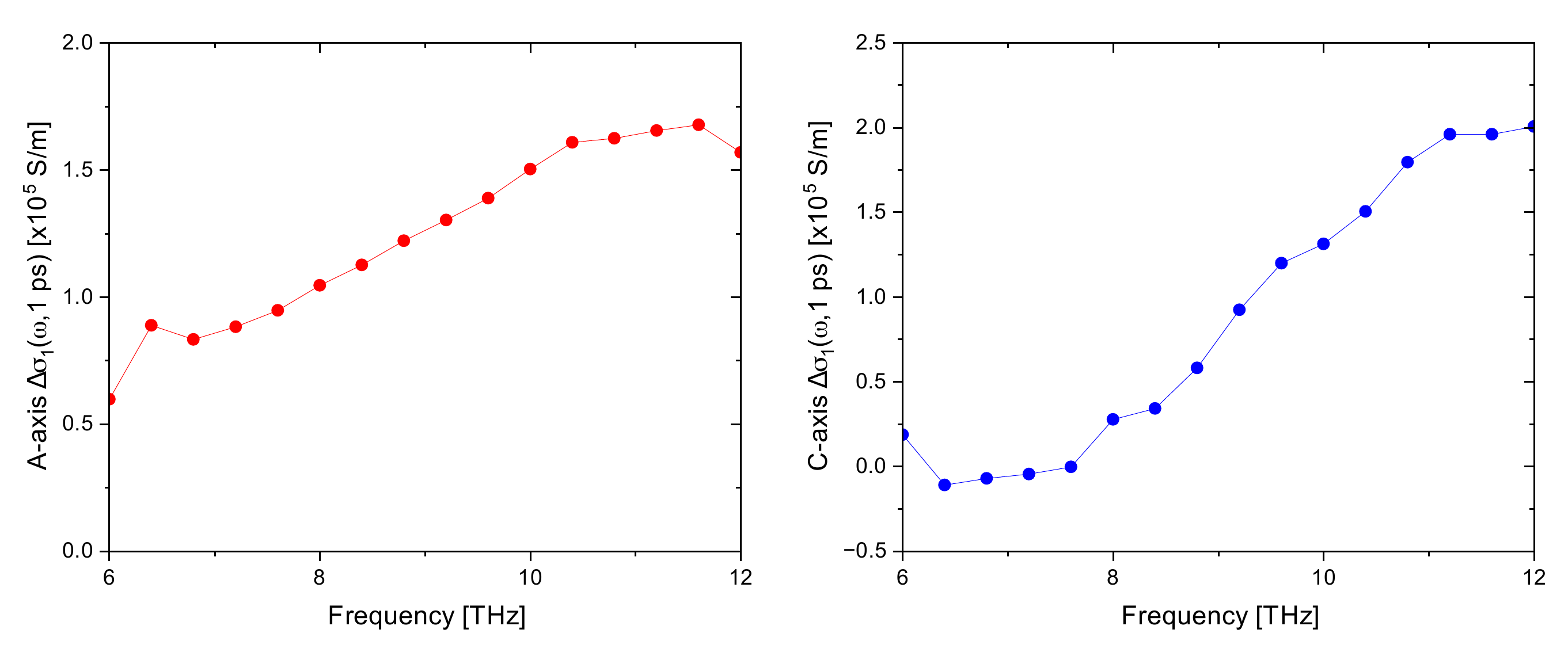}
    \includegraphics[width=0.85\columnwidth]{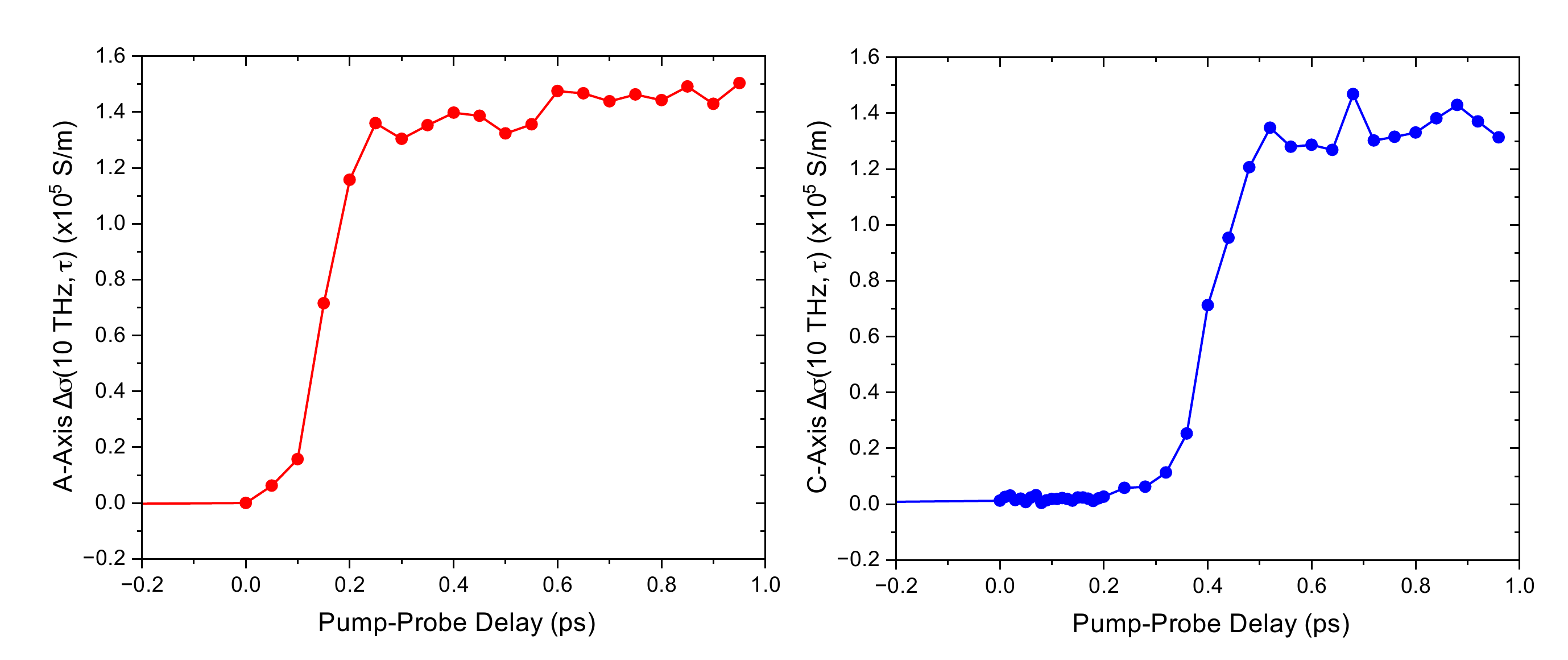}
    \caption{$\Delta \sigma_1$ spectra at $\tau =$ 1~ps (above) and dyanmics at $\omega / 2\pi =$ 10~THz for other SnSe crystal axes showing similar high-frequency change in conductivity. The a-axis is the interlayer stacking axis and the c-axis is the accordion axes, as shown in Figure 1. Excitation fluence was 9.5 mJ/cm$^2$. These results support the isotropic nature of the measured phase change. The relative delay of the c-axis dynamics is due to changes in the crystal surface position.}
    \label{sfig2}
\end{figure}

\clearpage

There is a possibility for artifact in the recovered conductivity spectrum for early pump-probe delays and small scattering times, as explained in Ref. [42]. To ensure these artifacts are not present in the presented datasets, the deviation was calculated and overlayed with the measured data, showing that these artifacts are not present. These artifacts are not of concern in SnSe due to carrier momentum relaxation times being extremely short. Results are displayed in Supplemental Fig. \ref{sfig3}.

\begin{figure}[t]
    \centering
    \includegraphics[width=0.7\columnwidth,trim={0cm 23cm 0cm 0cm},clip]{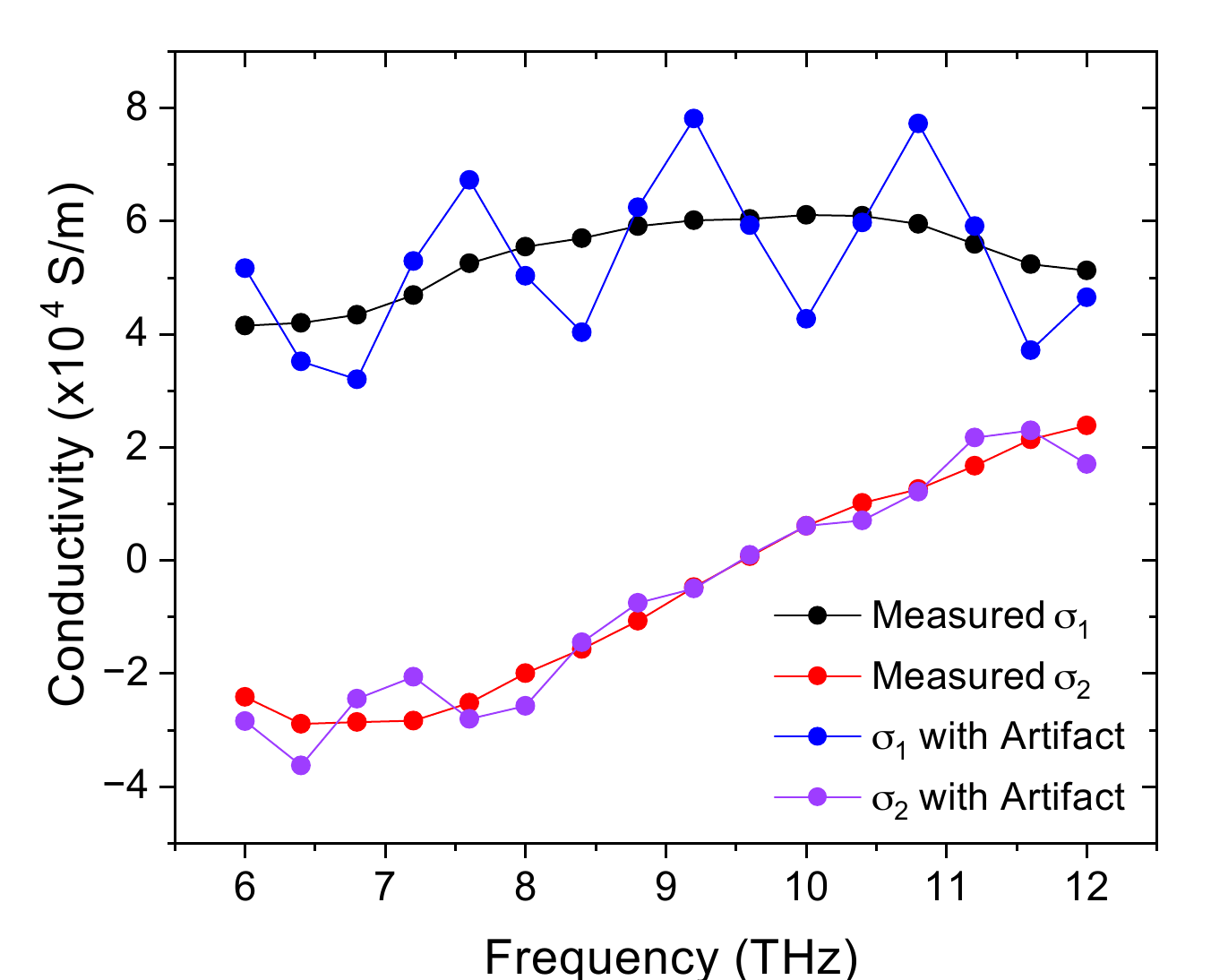}
    \caption{Calculated artifact signal superimposed over real conductivity data showing obvious oscillations that are not in any of the measured spectra. When parameters accurate to the measurements and material system are used, no significant deviations are expected.}
    \label{sfig3}
\end{figure}

\clearpage

\begin{figure}[h]
    \centering
    \includegraphics[width=0.45\columnwidth]{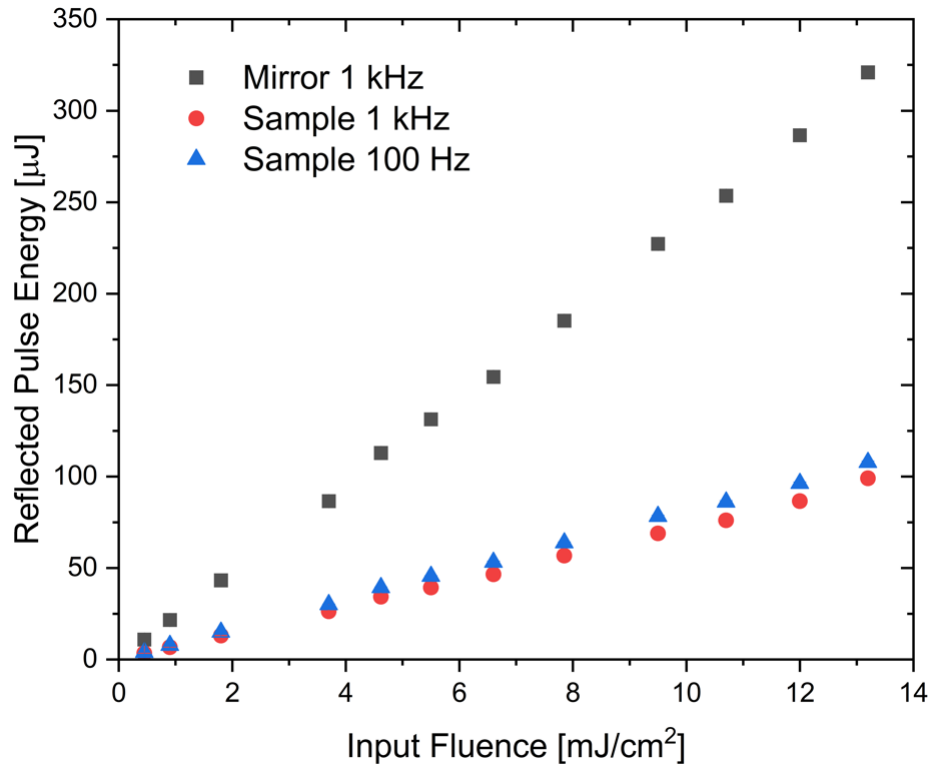}
    \caption{Fluence-dependent optical reflectivity measurements. Linear reflectivity change with input fluence shows there is no absorption saturation in the fluence range used for this study.}
    \label{sfig5}
\end{figure}

\begin{figure}[h]
    \centering
    \includegraphics[width=0.55\columnwidth]{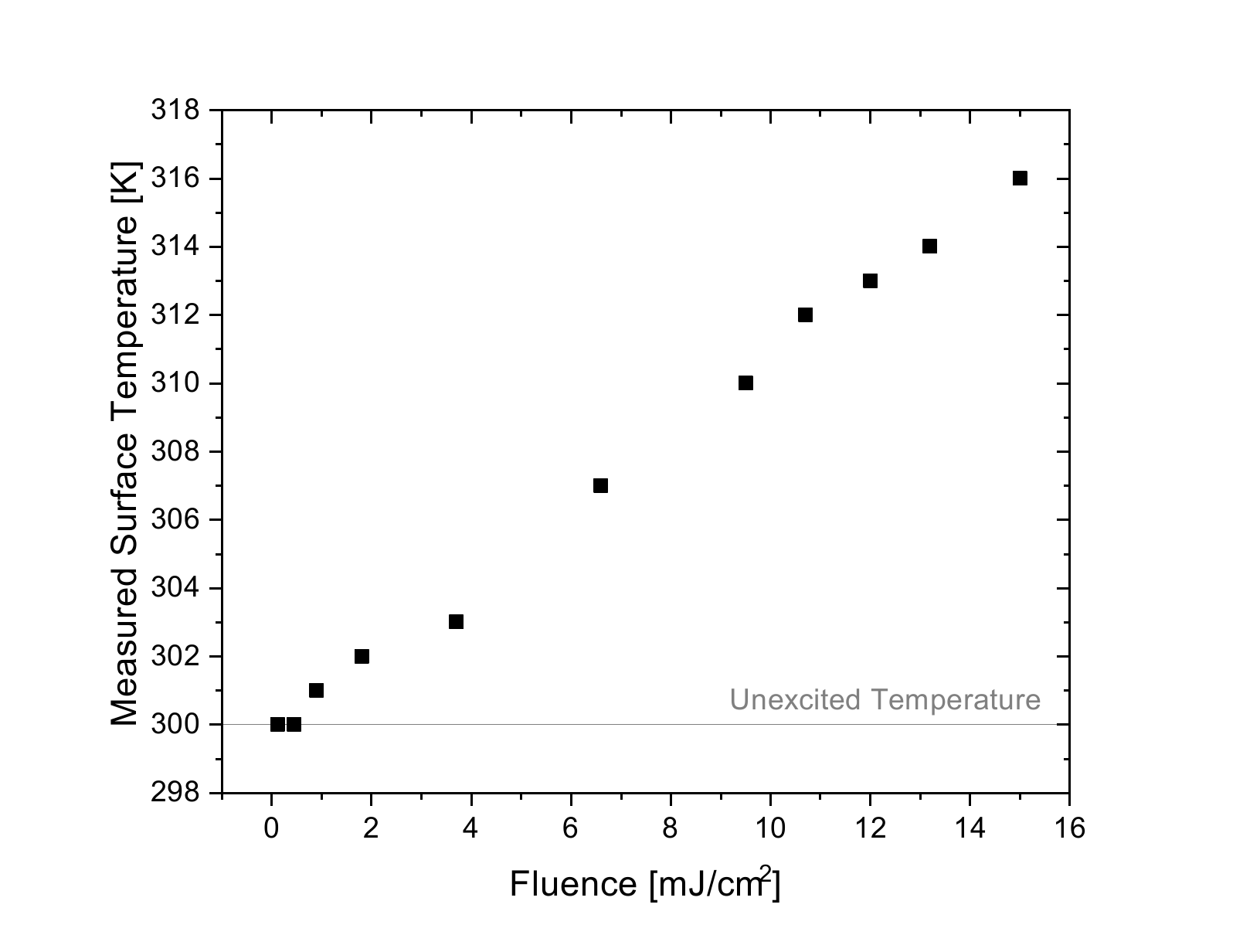}
    \caption{Sample surface temperature measurements as a function of incident pump fluence collected via thermal camera. Surface heating did not reach the level where thermally-assisted phase transitions or different carrier scattering rates will be activated.}
    \label{sfig6}
\end{figure}

Measurements of the fluence-dependent reflectivity (Supplemetal Fig. \ref{sfig5}) and surface temperature (Supplemental Fig. \ref{sfig6}) show no changes to the pump penetration depth and photoexcited lattice temperature, respectively. The 800~nm reflectivity shows no deviation from linear dependence in the fluence range investigated, which would signal a saturation of the photoexcited charge density. The maximal surface temperature of 16~K, measured with an infrared laser thermometer, at the highest incident optical fluence shows that no significant lattice temperature change is accumulated after repeated incident excitation pulses.

\clearpage

2D Re[$\Delta \sigma(\omega,\tau)$] and Im[$\Delta \sigma(\omega,\tau)$] maps for additional fluences are shown in Supplemental Fig. \ref{sfig7}. Blueshifting of the peak in the real part and zero crossing of the imaginary part are seen in fluences up to 6.6 mJ/cm$^2$, where the peak and zero crossing move to higher frequencies above the bandwidth for the experiment. The dynamics of the conductivity enhancement are also clearly more rapid in the higher fluence maps.

\begin{figure}[h]
    \centering
    \includegraphics[width=0.6\columnwidth]{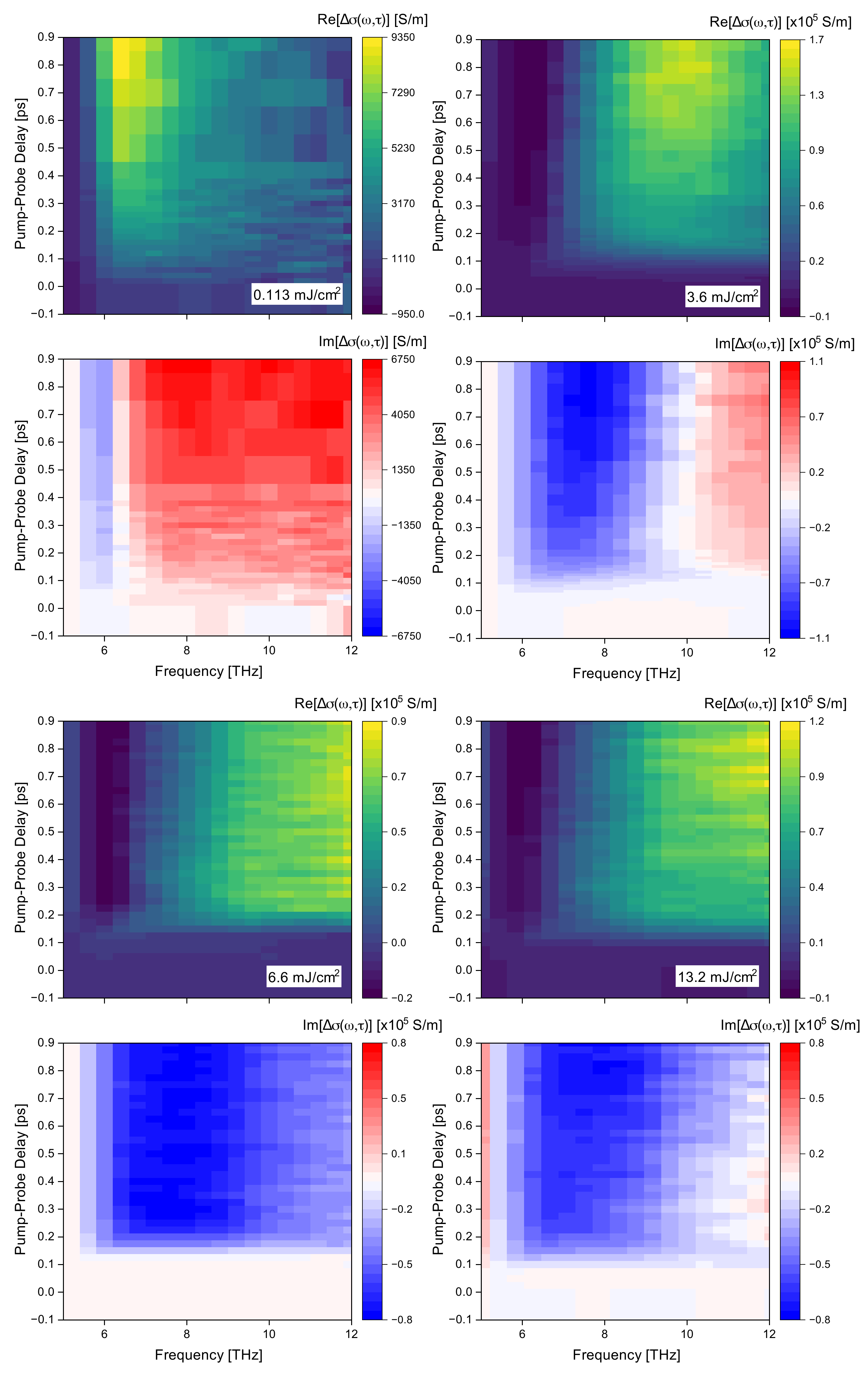}
    \caption{2D maps showing $\Delta \sigma_1 (\omega,\tau)$ and $\Delta \sigma_2 (\omega,\tau)$ for additional fluences. Note that the peak in  $\Delta \sigma_1$ and the zero crossing in $\Delta \sigma_2$ are linked as the pump-induced response blueshifts and broadens with increasing fluence.}
    \label{sfig7}
\end{figure}

\clearpage




The changes seen in the 2D maps are reflected in the spectra and dynamics cuts, with fluences in addition to those shown in Figure 2 shown in Supplemental Fig. \ref{sfig8}.

\begin{figure}[h]
    \centering
    \includegraphics[width=0.95\columnwidth]{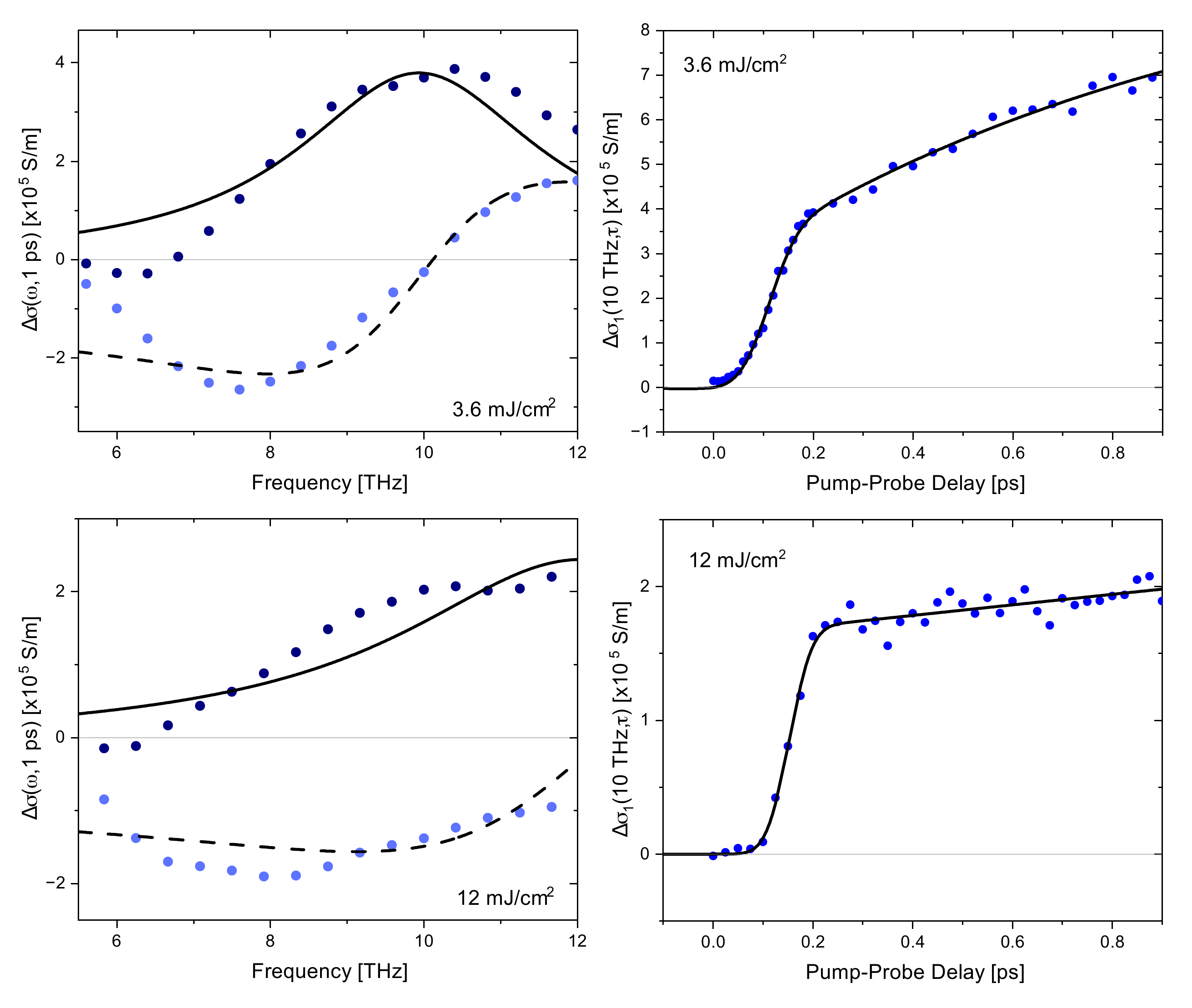}
    \caption{Lorentzian spectrum fits (left) and rate equation dynamics fits (right) for additional fluences 3.6 and 12.0 mJ/cm$^2$. Refer to Figure 2 for explanations.}
    \label{sfig8}
\end{figure}

\clearpage






Tracking of the long-time pump-induced change in reflectivity was achieved by measuring the change in the peak THz field at various times after pump arrival, with results shown in Supplemental Fig. \ref{sfig10}. The $\Delta R / R_0 (\tau)$ signal up to $\tau = $ 500~ps was recorded, limited by the length of the pump line mechanical delay stage. The time constant of the photoinduced change decay increases when approaching $F_c$, but quantitative analysis of these responses is difficult considering reflectivity changes are occurring on top of the already high static reflectivity in the reststrahlen band between 3 and 5~THz, shown in Supplemental Fig. \ref{sfig4}. This effect is avoided in the fluence-dependent 2D maps presented in the main text since data at probe frequencies above the high-reflectivity region can be isolated when the full THz electric field is sampled instead of only the peak field.

\begin{figure}[h]
    \centering
    \includegraphics[width=0.8\columnwidth]{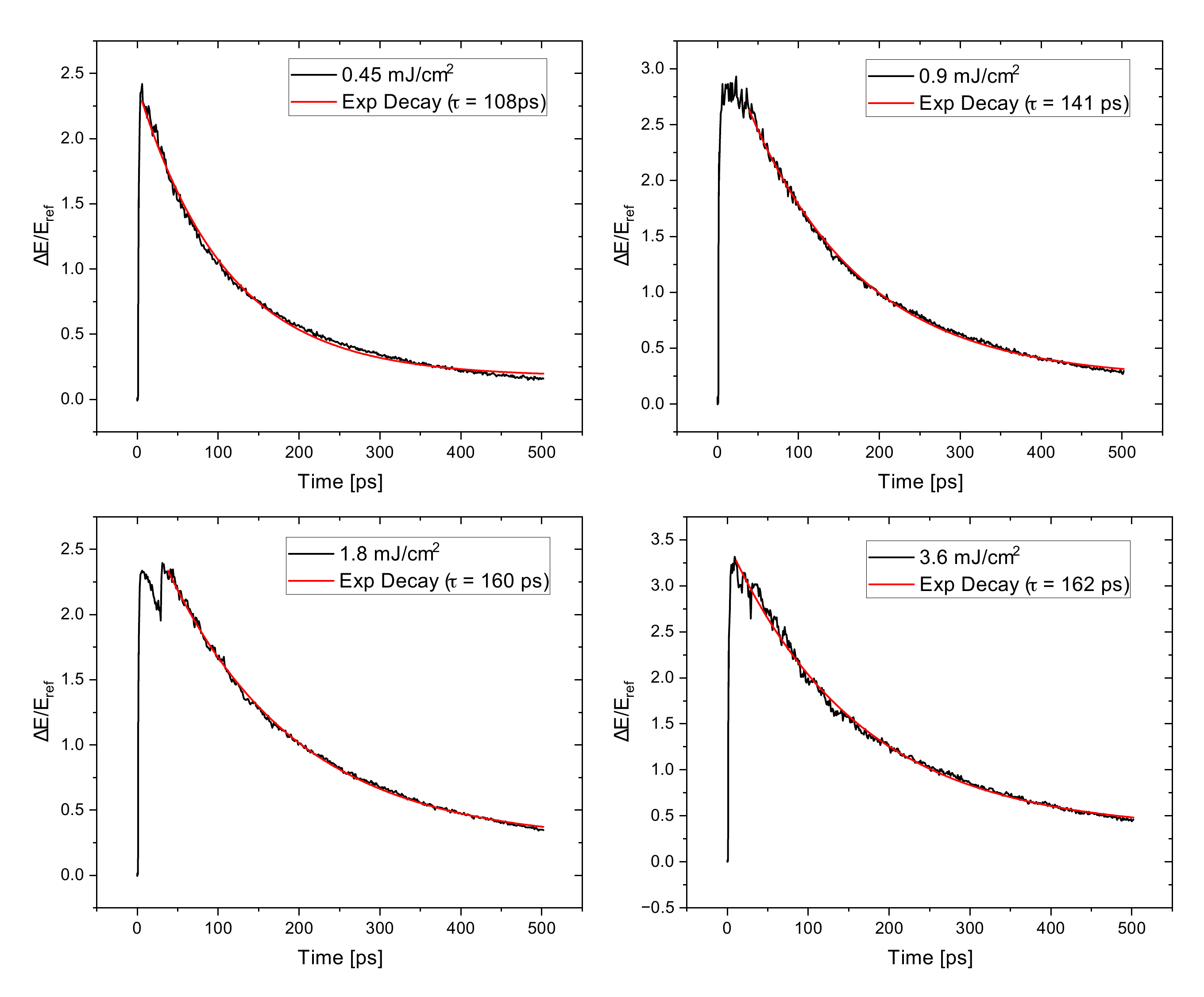}
    \caption{One-dimensional pump response decay curves, tracking changes in the THz peak field as a function of pump-probe time delay $\tau$. Exponential fits show that the response decays over 100-200~ps, in-line with previous reports of interband recombination.}
    \label{sfig10}
\end{figure}

\clearpage

Additional spectra at $\tau = 51$~ps, shown in Supplemental Fig. \ref{sfig11}, were taken at select fluences to track the change in the photoexcited conductivity spectrum at times greater than the $\tau = 0-1$~ps region investigated in detail at all fluences. At fluences above $F_c$ the enhancement in $\Delta \sigma_1 (\omega,\tau)$ is seen to redshift between $\tau = 1$~ ps and $\tau = 51$~ps, matching the relaxation seen in the peak electric field reflectivity decay. This redshift of the spectra signals carrier relaxation, as the long time spectra at these higher fluences resemble the spectra obtained at lower fleunces and therefore smaller excitation densities.

\begin{figure}[h]
    \centering
    \includegraphics[width=0.8\columnwidth]{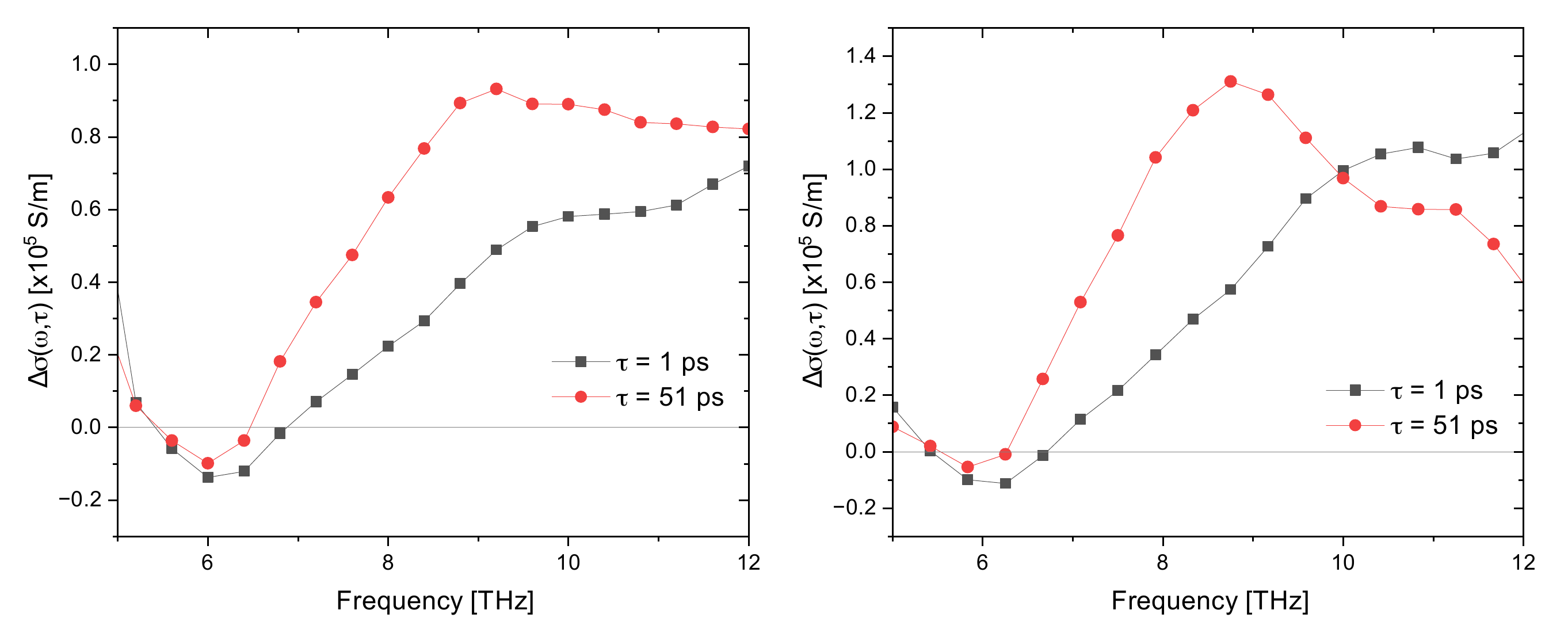}
    \caption{$\Delta \sigma_1$ spectra at $\tau =$ 1~ps and $\tau =$ 51~ps showing the long-time behavior past the 1~ps shown in the 2D maps. Fluences are 6.6 mJ/cm$^2$ (left) and 13.2 mJ/cm$^2$ (right). The 51~ps spectra show features present in the lower fluence data, signalling that the blueshift and broadening seen above the fluence threshold is indeed a density driven effect which relaxes as carriers relax and recombine over time.}
    \label{sfig11}
\end{figure}


\bibliography{SnSe_Mar23_Abbrv.bib}